\documentclass[peerreview,conference]{IEEEtran}
\ifCLASSINFOpdf
\else
\fi

\usepackage{fancyhdr}
\usepackage{hyperref}

\usepackage[textsize=small]{todonotes}
\usepackage{siunitx}
\usepackage{subcaption}
\usepackage{amsmath}
\usepackage{graphicx}
\usepackage{cite}
\usepackage{xspace}
\usepackage{multirow}
\usepackage{setspace}

\newcommand{\ignore}[1]{}

\newcommand{\microsubmissionnumber}{208}
\fancypagestyle{firstpage}{
  \fancyhf{}
\setlength{\headheight}{10pt}

  \fancyhead[C]{\normalsize{MICRO 2018 Submission
      \textbf{\#\microsubmissionnumber} -- Confidential Draft -- Do NOT Distribute!!}}
  \pagenumbering{arabic}
}

\title{Guidelines for Submission to MICRO 2018}
\begin{document}
%
\title{SatIn: Hardware for Boolean Satisfiability Inference}

\author{\IEEEauthorblockN{Chenzhuo Zhu, Alexander C. Rucker, Yawen Wang, and William J. Dally}

\IEEEauthorblockA{Stanford University\\Stanford, CA 94305\\Email: \{czhu95, acrucker, yawenw, dally\}@stanford.edu}
}

\maketitle
\setcounter{page}{1}

\newcommand{\minisat}{MiniSAT\xspace}
\newcommand{\glucose}{Glucose\xspace}
\newcommand{\todoval}{XXX\todo{value}\xspace}

\begin{abstract}
    This paper describes SatIn, a hardware accelerator for determining boolean satisfiability (SAT)---an important problem in many domains including verification, security analysis, and planning. 
    SatIn is based on a distributed associative array which performs short, atomic operations that can be composed into high level operations. 
    To overcome scaling limitations imposed by wire delay, we extended the algorithms used in software solvers to function efficiently on a distributed set of nodes communicating with message passing.
    A cycle-level simulation on real benchmarks shows that SatIn achieves an average 72x speedup against \glucose \cite{glucose}, the winner of 2016 SAT competition, with the potential to achieve a 113x speedup using two contexts.
    To quantify SatIn's physical requirements, we placed and routed a single clause using the Synopsys \SI{32}{nm} educational development kit.
    We were able to meet a \SI{1}{ns} cycle constraint with our target clause fitting in \SI{4867}{\mu m^2} and consuming \SI{63.8}{\mu W} of dynamic power; with a network, this corresponds to 100k clauses consuming \SI{8.3}{W} of dynamic power (not including leakage or global clock power) in a \SI{500}{mm^2} \SI{32}{nm} chip.
\end{abstract}

\section{Introduction}
Boolean satisfiability (SAT) is an NP-complete problem that forms the core of many practical problems, including planning and formal design verification \cite{kautz1992planning, biere1999symbolic, bryant2002modeling}. SAT attempts to find an assignment of boolean values to variables that results in a given formula being true; it is also possible to reduce many problems in NP to SAT in polynomial time \cite{complexity}. Solvers are under continuous development, with an annual SAT competition highlighting the latest advances \cite{sat16}. Because SAT can easily be optimized, many practical problems are translated into SAT instead of being solved directly \cite{een2006translating, practical}. 

Prior work has been done to accelerate SAT using custom hardware \cite{davis2008practical, thong2013fpga}. However, both the overall speedup against state of the art software solvers and the maximum problem size are very limited. This is due to two reasons: traditional architectures store data (clauses) separately from processing elements, making it expensive to traverse all clauses once SAT problems get large enough, and most proposed hardware accelerators are only able to accelerate a single part of modern SAT solving algorithms. Because the remaining part still takes 10\% of the total runtime, the overall speedup is limited to 10x.

\begin{figure}[t]
    \centering
    \vspace{-1.5in}
    \hspace{-1.7in}
    \includegraphics[width=1.1\textwidth]{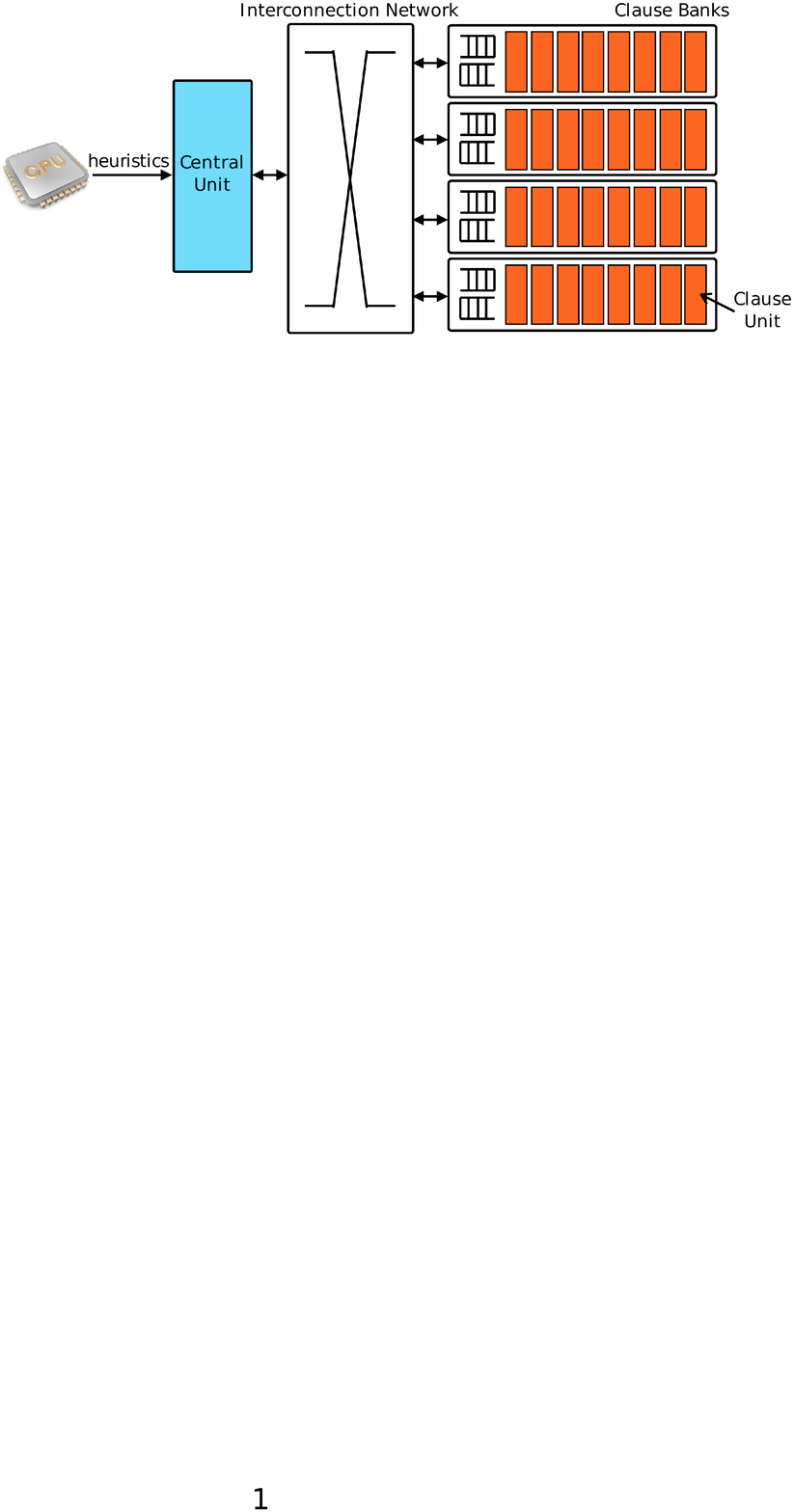}
    \vspace{-6.5in}
    \caption{Top level architecture of SatIn. Central unit and clause banks are connected through an interconnection network.}
    \label{intro:top}
\end{figure}

To address these problems, we build custom hardware for clause storage and basic clause operations to eliminate clause data movement, even for large SAT problems. Instead of moving clause data, we deliver variable assignments to each clause through an interconnection network (Figure~\ref{intro:top}). The latency of the interconnection network is hidden by allowing asynchronous implications. Furthermore, our clause units are reused for multiple computationally intensive parts of modern SAT algorithms, minimizing CPU load while leaving flexibility for heuristic tuning. With more than 99\% of the baseline SAT algorithm accelerated by our hardware, we achieve an average 72x speedup with a single context against the 2016 SAT competition winner \cite{glucose}. Our design is easy to expand for very large SAT problems, either with a larger chip or by using multiple chips.

To make our accelerator distributed, and therefore scalable, we introduce several modifications to single threaded SAT algorithms. These allow key steps of the algorithm, such as constraint propagation and clause learning, to run in parallel while being correct.
To evaluate the speedup from our accelerator, we use a custom cycle-approximate simulator to run several problems from the 2016 SAT competition \cite{sat16}. We compare the speedup against \glucose, using the same decision heuristics and learned clauses to ensure a fair comparison \cite{glucose}. We extrapolate the results from this speedup to identify an ideal speedup for adding a second, independent thread.
We then evaluate the cycle time, area, and power of our accelerator by using the Synopsys \SI{32}{nm} educational development kit to place and route a clause unit and router and analytically model the network links. This shows that, in a \SI{500}{mm^2} \SI{32}{nm} die, we can fit up to 100,000 clauses with a total power consumption of \SI{8.25}{W}. Scaling this to a modern 16nm process should allow upwards of 400,000 clauses on a single die of comparable size.

This paper makes the following contributions:
\begin{enumerate}
\item We analyse the important characteristics of modern SAT problems and use these characteristics to motivate our hardware design,
\item We develop a distributed SAT algorithm, optimized for minimal synchronization overhead that permits the use of
existing heuristics from state-of-the art software SAT solvers,
\item We develop the architecture of SatIn, a SAT accelerator that uses a distributed architecture of many simple clause units connected by a NoC to accelerate the bulk of the SAT problem, and
\item We evaluate SatIn on problems from the 2016 SAT competition.  Our evaluation includes a performance comparison against Glucose, the software program that won the 2016 competition.  We also include area and power numbers from implementing SatIn in a 32nm technology through place and route.
\end{enumerate}
We start by describing modern SAT algorithms and our modifications in Section~\ref{sec:alg}. System architecture and clause architecture of the accelerator are described in detail in Section~\ref{sec:sys} and Section~\ref{sec:cla}. Section~\ref{sec:meth} describes our evaluation methodology and is followed by experimental results in Section~\ref{sec:res}. We discuss the related work in Section~\ref{sec:rel} and conclude in Section~\ref{sec:concl}.

\section{SAT Algorithms}
\label{sec:alg}

For this work, we focus on SAT problems that are expressed in Conjunctive Normal Form (CNF). This is a product of sums representation: a list of \emph{clauses}, where each clause is a list of \emph{literals}. Each literal is an assignment of a \emph{variable} to a \emph{polarity}, either negated or not. For the problem to be satisfied, every clause must be satisfied; for a clause to be satisfied, at least one literal in the clause must be true. CNF allows the solver to reason about the formula more efficiently, while still being a reasonable target for translating practical problems. In this section, we will first describe common algorithms used in modern software SAT solvers and then propose our modification to each algorithm for distributed execution.

\subsection{Software Algorithms}

Most deterministic SAT algorithms today are based on the original algorithm by Davis, Putnam, Logemann, and Loveland (DPLL), which uses a tree search to explore the set of potential solutions \cite{dpll}. Each node in a tree represents the decision of a single variable to be either true or false, and the algorithm explores the tree, turning back as soon as it finds that a particular decision precludes a satisfying assignment. Over the past decades, extensions to DPLL and smart heuristics have been proposed to allow faster SAT solvers.

\subsubsection{Boolean Constraint Propagation}
The most important step in a DPLL-based SAT solver is Boolean Constraint Propagation (BCP). When a CNF clause has all but one literal decided, and no literals are decided as true, it is referred to as a \emph{unit} clause. Unit clauses must have their remaining literal decided as true to have the problem satisfied, so the literal is decided to be true and then applied to all other clauses. For example, consider a problem with clause $x_0 \lor x_1' \lor x_2'$ and the existing decisions $x_0'\quad(x_0\rightarrow 0)$ and $x_1\quad(x_1\rightarrow 1).$ For this clause to be satisfied, $x_2$ must be 0, and this logical \emph{implication} is broadcast to the remaining clauses. However, the implication $x_2'$ depends on $x_0', x_1.$ Therefore, if either $x_0'$ or $x_1$ is undone, then $x_2'$ must also be undone.
Software solvers are able to efficiently implement this algorithm using the two-watched literal scheme introduced by Chaff \cite{chaff}. Instead of checking all literals in all clauses, these solvers check only two, which is sufficient to ensure that a clause has not become unit. 
Once a watched literal is decided, the clause is satisfied, unit (if all unwatched literals are decided), or another literal is swapped in to be watched. This is increasingly efficient for longer clauses, because most implications do not need to be handled at all clauses that they impact.

However, it is possible that the logical results obtained from BCP may produce conflicting implications. Consider the following clauses: 
$$ x_0 \lor x_1' \lor x_2' $$
$$ x_0 \lor x_1' \lor x_2. $$
If $x_0',x_1$ is decided, then the two clauses will attempt to propagate $x_2$ and $x_2'.$ Because a variable cannot logically be in two states, this results in a \emph{conflict}, and either $x_0$ or $x_1'$ must be true for a satisfying assignment. 
The problem is not satisfiable if there are no remaining branches of the search space to explore after a conflict. 

\subsubsection{Decision Heuristics}
A literal's decision may trigger a series of implications. If none of these implications result in conflicts and all clauses are not yet satisfied, another literal needs to be decided to continue the search. To associate implications to their triggering decision, a \emph{decision level} is kept globally and incremented when a literal is decided. 
SAT solvers rely on decision heuristics like the Variable State Independent Decaying Sum (VSIDS) heuristic, first introduced by the Chaff solver \cite{chaff}. VSIDS keeps a periodically decaying sum for each literal, which is incremented whenever that literal is part of a learned clause. When the solver must make a decision, it chooses the undecided literal with the highest count. In essence, VSIDS facilitates search progress by attempting to satisfy recent conflict clauses first.  

\subsubsection{Conflict-Driven Clause Learning}
The GRASP solver improves upon the original DPLL procedure by learning new clauses whenever conflicts occur \cite{grasp}. The new clauses are added to the database, and are used to backtrack farther up the search tree after resolving a conflict. These learned clauses are used later to avoid reaching the same conflict in a different branch of the search space. 

A conflict is detected when both literals for a variable are implied as true. The last literal assignment that led to both implications is defined as the first Unique Implication Point (1-UIP). The new clause is the negated 1-UIP literal and the negation of any literals from earlier decision levels that led, directly or indirectly, to the conflicting implication. Instead of backtracking to the previous decision level, the solver returns to the latest decision level where the learned clause is unit, generating an implication that sets the 1-UIP literal to true. BCP is then performed as usual. 

For example, consider the three clauses $$x_0\lor x_1\lor x_2,$$ $$x_1\lor x_3',$$ $$x_2'\lor x_3.$$ Deciding $x_0'$ (decision level 1) triggers no implications. Deciding $x_1'$ (decision level 2) triggers the implication of $x_3'$. $x_0'$ and $x_1'$ together imply $x_2$, which then triggers a conflicting implication of $x_3$. In this example, the 1-UIP literal of the conflict pair is $x_1',$ and $x_0'$ is the only assignment from earlier decision levels contributing to this conflict. 
$x_1\lor x_0$ is added to the database, and the solver backtracks to the point immediately after the decision of $x_0'$, which then generates the implication of $x_1$.  

\subsubsection{Clause Strengthening}
Shorter learned clauses are desirable because they require less space and become unit more quickly. \minisat introduced a clause strengthening procedure to shorten learned clauses by recursively applying a logical simplification \cite{minisat}. Strengthening identifies variables in the learned clause that are directly or indirectly implied by other literals in other learned clause and removes them.   
In addition, minimization with binary resolution is applied on the asserting literal in the learned clause to further reduce learned clause length \cite{luo2017effective}.

\subsubsection{Clause Deletion and Restarts}
Another important heuristic used by modern SAT solvers is the clause deletion heuristic: although the solver learns many clauses during its execution, not all of them prove to be equally useful. An effective deletion heuristic allows a solver to occasionally prune the database of learned clauses \cite{chaff, glucose}. 
In \glucose, each learned clause is associated with a Literal Blocks Distance (LBD) score which is the number of different decision levels of literals contained in that clause. Clauses with lower LBD are more desirable because their literals can be assigned with fewer decisions. 
\glucose aggressively cleans out its database by removing the half of the learned clauses with the highest LBD scores after a certain number of conflicts. 

Additionally, solvers such as \glucose periodically restart while keeping their learned clauses and heuristics. This is important to avoid remaining ``stuck'' in an unproductive branch of the search tree, which can happen if an early decision precludes a satisfying assignment \cite{grasp, chaff}.

\subsubsection{Parallelism}
Several methods of parallelism have been proposed and implemented for software SAT solvers. The simplest case is that the problem is divided into sub-problems by assigning some combination of fixed variables to processes: for example, work could be distributed among eight processes by assigning $x_0,x_1,x_2$ to the first process, $x_0,x_1,x_2'$ to the second, and so forth. However, this ignores the fact that a satisfying solution, if one exists, may only be found in a subset of branches and others might be trivially unsatisfiable. 
Therefore, this static work distribution may result in some processes finishing significantly earlier than others, and these solvers must partition the problem space into potentially millions of sub-problems and dynamically share the work \cite{cube}.
Another possibility involves using multiple processes, each attempting to solve the same problem; this is the approach that \glucose and many other parallel solvers use \cite{syrup}. The processes are initialized with different random number generator states, and potentially different heuristics, before racing to completion. While the processes are running, they periodically select and share learned clauses. Sharing learned clauses allows one process to learn from conflicts another process has encountered, and the use of different heuristics means that at least one thread of the solver is likely to have heuristics that are well-adapted to the problem.

\subsection{Algorithm Modifications}
Modern software solvers are designed to be efficient running on high performance CPUs. However, they are not suitable for direct hardware implementation. In order to make an accelerator run efficiently, we analyzed major parts of modern SAT solvers and modified them to use the parallelism present in hardware.

\begin{figure}
    \centering
    \vspace{-.5in}
    \includegraphics[width=.7\textwidth]{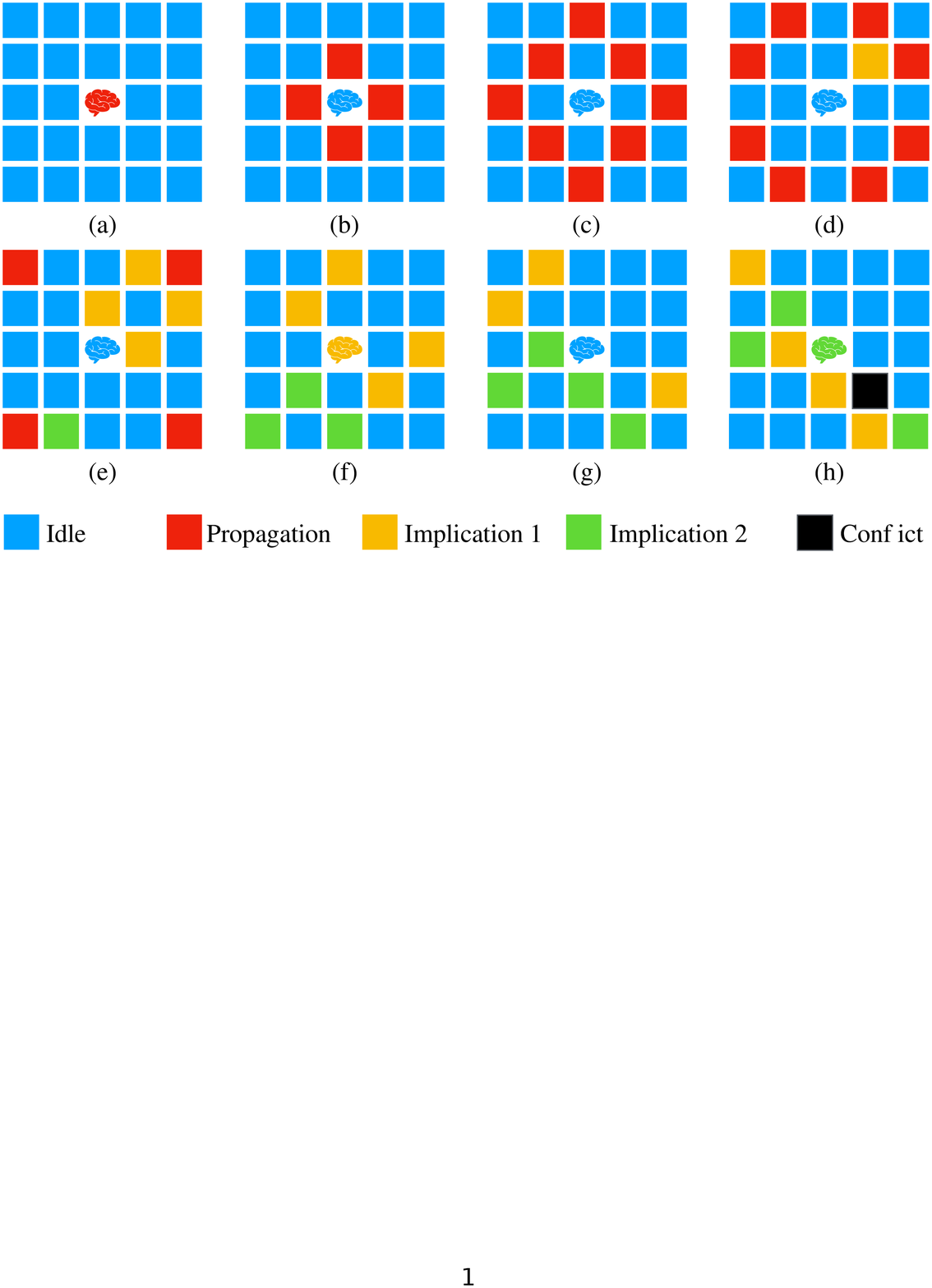}
    \vspace{-3in}
    \caption{Distributed boolean constraint propagation algorithm. The propagation starts from the central unit (a) and is broadcast to all clause units in the network. Implications are triggered on the path of the propagation (d) and (e). Conflicts can be detected locally in the conflict clause (h).}
    \label{alg:prop}
\end{figure}

\subsubsection{Distributed Boolean Constraint Propagation}
\label{subsec:dbcp}
With dedicated hardware support for all clauses in a SAT problem, one could perform a propagation on each clause simultaneously in a single cycle. However, it takes much longer for the propagated literal to arrive due to wire delay. Instead of handling propagations serially, we propose to run multiple propagations asynchronously and synchronize when there are no more propagations or is a conflict. 
Propagation latency can thus be hidden when there are enough ongoing implications to keep all clauses busy. An illustration of distributed BCP is shown in Figure~\ref{alg:prop}.

With distributed BCP, clauses detect conflicts locally and broadcast a global stop signal, eliminating a round trip through the central unit. This typically happens when a clause receives a propagation message with a variable marked as decided to the opposite polarity from a variable already locally decided. 
When variable polarities differ after a conflict, the resulting implications are logically meaningless. Therefore, for effective clause learning, the accelerator must identify the earliest conflict pair. To address this problem, we introduce a logical \emph{implication level} as an indicator of implication dependences and determine the conflict pair used for clause learning at a stand-alone central unit; this is similar to the use of logical clocks for providing a happens-before relationship in distributed systems \cite{lamport}.

The implication level is a number associated with each BCP message. Each propagation variable comes with an implication level $l_p,$ and we maintain a shared implication level $l_i$ for each clause bank. After each propagation, the shared implication level is refreshed as $\max(l_p, l_i),$ and outgoing propagations are sent with level $l_i+1.$ The implication level of an implied variable $x$ is therefore greater than that of all variables $x$ depends on, and the central unit can then easily maintain a single implication trail. correct conflict clauses for learning are then the conflicting pair with the lowest implication level, which is guaranteed not to be the result of a prior conflict.

Another function of implication levels is to provide unique reason clauses in distributed BCP. Software solvers like \glucose recursively track reason clauses for both clause learning and clause strengthening, and have exactly one clause as the reason for each implied variable. Uniqueness of reason clauses is guaranteed when variables are propagated sequentially because each variable can only be propagated once. For distributed BCP, two clauses at different corners of the chip can trigger the same implication for one variable. In this case, we depend on the central unit to specify a unique reason clause for the implication and mark all other clauses as not reason clauses. This is done by discarding all implications except the one with the lowest level; logically, this holds because $x$ implies $x$ and the higher-numbered implications are redundant.

\subsubsection{Distributed Clause Learning}
In \minisat and \glucose, clause learning is performed by recursively tracking reason clauses starting from the conflict pair. Literals implied in previous decision levels are negated and added to the learned clause, and literals from the current decision level are reasoned further. The tracking stops when a single unreasoned literal is found as the cut between the last decision and the conflict, or Unique Implication Point (UIP). The solver then jumps back to the latest level that will leave one literal in the learned clause unset, and sets it to satisfy the learned clause and start a new round of implications \cite{grasp}.

As described in Section~\ref{subsec:dbcp}, our distributed BCP implementation guarantees unique reason clauses, which are indicated by setting a bit within the clause. With our clause units, clause learning can be performed by our accelerator in the reverse order of BCP: after a conflict pair is determined by the central unit, a message is sent to each reason clause for the conflicting implications, to query about variables that triggered the implications. Each reason clause, upon receiving the query that matches its implication, generates a query message for each of other variables in the clause implied after the current decision level. All messages are also received by the central unit to build the learned clause.
Distributed clause learning reuses most of the hardware resources for BCP and requires only minimal storage in the central unit for the learned clause. For complicated problems, the number of queries on a single conflict easily grows to over a thousand, and the average number of cycles spent on each query is very close to 1.

\subsubsection{Backtracking}
When conflict occurs after a decision and a learned clause is generated, the central unit finds a decision level to backtrack to and cancels all variable decisions and implications after the backtrack level. Software solvers keep a global array of variable assignments to make backtracking cheap. 
However, we keep variable assignments locally in each clause unit, requiring that cancelled variables be broadcast to all clause units for backtracking. To reduce this penalty, we maintain a \emph{current} bit per literal in all clause units indicating the variable was decided or implied in the latest decision level. When there is a conflict, we can cancel all variables in the current decision level with one message instead of one per variable, but we still need to cancel variables decided earlier than current decision level sequentially. As most cancelled variables are in the current level, adding the current bit saves around 90\% of backtracking cycles.

\subsubsection{Distributed Clause Strengthening}
We implemented both clause strengthening mechanisms in a distributed style. We use an additional bit per literal to represent whether the literal is in the learned clause before strengthening. Once we learn a clause, the central unit sends messages to all clause units to mark all literals in the newly learned clause as learned. If a reason clause has all but the originally-implied literal marked as learned, it will generate a new message indicating its reason literal can be removed from the learned clause. Furthermore, this message will also mark the implication literal as learned in other reason clauses to enable removal of indirectly-implied literals from the learned clause.
The accelerator delays clause strengthening and performs it in parallel with implication, an approach introduced by Wieringa \cite{solverreducer}. Clause strengthening must only be completed before the next backtracking to ensure consistent variable assignments. This is because the backtracking level after the conflict can be calculated without strengthening, and the learned clause will not produce any implications until the solver backtracks beyond this level.


\subsubsection{Multiple Contexts}
\label{alg:context}
Because any effective hardware SAT solver will periodically need to synchronize and engage software running on a general purpose core, it is likely that there will be idle periods between bursts of propagations. Although these can be kept as short as possible, we still want to ensure that we are making optimal use of the hardware by exploiting multi-threaded parallelism, where logically independent threads explore different parts of the search space while sharing learned clauses. However, while \glucose is only able to share a few clauses due to communication overhead, we are able to share \emph{all} clauses---by doing so, we avoid duplicating the literals in each clause, and only have to duplicate the status bits (an approximately 25\% increase in state).

\begin{figure*}
    \centering
    \vspace{-.5in}
    \includegraphics[height=3in]{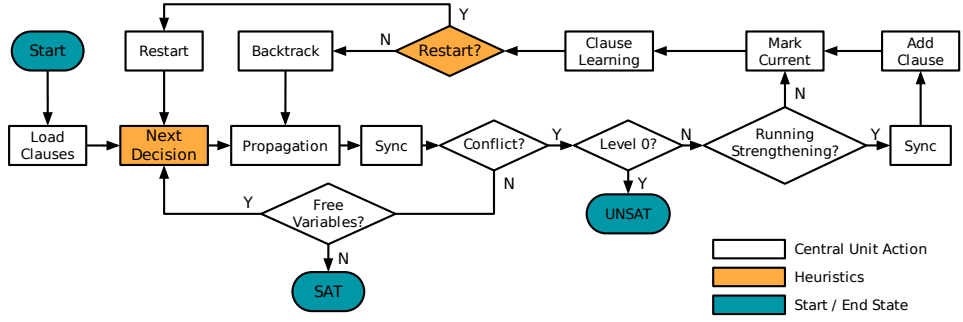}
    \vspace{-.5in}
    \caption{Control logic for central unit.}
    \label{sysarch:flow}
\end{figure*}

\section{System Architecture}
\label{sec:sys}
Our system consists of a large number of clauses and a central controller connected together via a network, as shown in Figure~\ref{intro:top}. To allow the clauses to be efficiently accessed, they are grouped into banks, where each bank shares hardware including the network router and sequencing logic to convert certain longer messages into the simpler commands that clauses are able to execute. 
The core of the central unit is a general purpose CPU executing the control flow shown in Figure~\ref{sysarch:flow}, which allows the architecture to be flexible in which heuristics it supports. Because heuristics are critical to obtaining good performance from a SAT solver and are still an active area of research and commercial differentiation, an accelerator must be adaptable to be competitive. The central unit also includes a received implication sorter, to provide a consistent serial view of the parallel implication execution by ordering received BCP messages by implication level.

\subsection{Network Design}

\subsubsection{Messages}
We designed a set of messages that enables communication among clause banks as well between clause banks and the central unit, which are combinations of the basic fields shown in Table~\ref{tb:field}, with a complete list of messages in Table~\ref{tb:message}. Unlike interconnection networks that use point to point routing, most of our messages are broadcast to all other routers in the network. Therefore, instead of using only network addresses, we use 3 routing option bits to allow: 
\begin{enumerate}
    \item routing back to the source clause bank, 
    \item broadcast, and 
    \item routing to the central unit. 
\end{enumerate}
Combinations of these routing options suffice for most routing, and a network address can still be used when the message is bound for a specific clause bank. With the field lengths specified in Table~\ref{tb:field}, we can support address up to $2^{20}$ clauses and variables. All messages except \texttt{AddClauseMessage} are short enough ($<64$ bits) to fit into one flit, which minimizes network latency and simplifies router design. \texttt{AddClauseMessage} is divided into several packets and reassembled at the destination clause bank.

\begin{table}
    \centering
    \begin{tabular}{l|ll}
                          Name & Abbrv & Length (bits) \\ \hline
                          Header            & H      & 6             \\
                          Network address   & N      & 10            \\
                          Clause address    & C      & 10            \\
                          Variable          & V      & 20            \\
                          Polarity          & P      & 1             \\
                          Implication level & I      & 14           \\
                          Extra bits		& E 	 & 1
    \end{tabular}
    \caption{Field types and lengths in a message.}
    \label{tb:field}
\end{table}

\begin{table*}
    \centering
    \begin{tabular}{l|llllllll|p{3in}}
        Name      & H & N & C & V & P & I & E & Bits & Description                                                     \\ \hline
        AddClause  & 1 & 1 & 1 & $\le$8 & $\le$8 & 0 & 0               & $\le$194                 & Add a clause to a given clause bank                             \\
        PropLit    & 1 & 1 & 1 & 1 & 1 & 1 & 1               & 62                  & Propagate one literal with an implication level                 \\
        CancelVar   & 1 & 0 & 0 & 1 & 0 & 0 & 0               & 25                  & Cancel a previous decision or implication of a variable         \\
        CompleteDL & 1 & 0 & 0 & 1 & 0 & 0 & 0               & 25                  & Mark the beginning of a new decision level                                \\
        Conflict   & 1 & 0 & 0 & 0 & 0 & 1 & 1               & 21                  & Bring a stop to all clause banks after a conflict               \\
        NotReason  & 1 & 1 & 1 & 0 & 0 & 0 & 0               & 25                  & Mark a clause as not a reason clause                        \\
        Reason     & 1 & 0 & 0 & 2 & 2 & 0 & 0               & 48                  & Query about the literals that triggered an implication          \\
        Strengthen & 1 & 0 & 0 & 1 & 1 & 0 & 0               & 27                  & Mark a literal as removable in the learned clause
    \end{tabular}
    \caption{The set of messages used in the network.}
    \label{tb:message}
\end{table*}

\subsubsection{Router Design}
\label{sysarch:router}
We make two major modifications to network routers compared with traditional interconnection networks: a simpler flow control mechanism and broadcast routing. Virtual channels are not necessary because all messages fit in a single flit. A credit-based flow-control mechanism with input buffers is used. Deterministic, dimension-order routing is used to broadcast messages: sending them first to all routers in the same row and then to all routers in each column. To reduce latency, a single-cycle router is employed.  Route calculation, traversal of wires to output multiplexers, and output arbitration are all performed in one cycle.  Flits begin traversing the output channel on the next cycle. We explored both mesh and flattened butterfly topologies, with the central unit placed at the center of the network to reduce latency.

\subsubsection{Idle Detection}
Another important function of our network is detecting global idle---the state when no messages are being processed or will be generated in the network for a given context. It is important to notify the central unit as soon as network becomes idle to reduce synchronization overhead and allow the central unit to broadcast the next decision. We monitor activity in each clause bank and router and then combine these local idle signals in an AND tree. Dedicated idle signal wires have lower overhead than routing idle signals as messages.

\begin{figure}
    \centering
    \includegraphics[width=.5\textwidth]{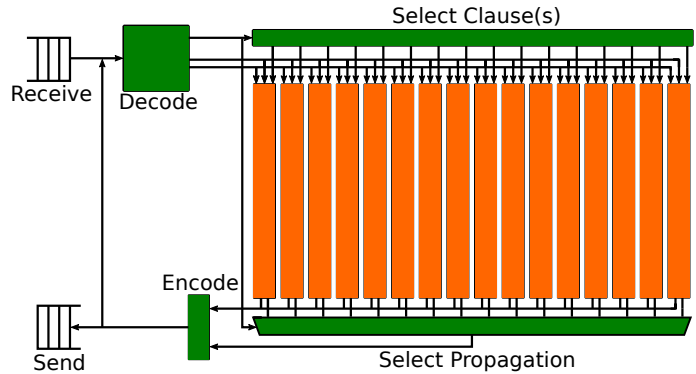}
    \caption{Clause bank architecture.}
    \label{sysarch:arch}
\end{figure}

\section{Clause Architecture}
\label{sec:cla}
Because each individual clause handles far more incoming propagation or backtracking messages than any other type of message, the primary purpose of the clause unit is to provide an extremely fast and efficient associative search mechanism. Clauses also need to be as small as possible: the largest problem that can be solved is limited by the number of clause units. Therefore, we offload as much complexity as possible into the central controller and a per-bank sequencing unit that communicates with the network. Clauses have no internal sequencing, and operate with a small set of inputs and outputs to allow integration into large banks. Because hardware does not provide good support for variable length clauses, we restrict problems to have fixed 8-variable clauses. Variables can be introduced to provide logical relationships between split clauses, so this does not limit the generality of our accelerator. For example, consider the clause $x_0 \lor x_1 \lor x_2 \lor x_3.$ By adding a variable $c_0,$ the problem can be rewritten as $(x_0 \lor x_1 \lor c_0) \land (c_0' \lor x_2 \lor x_3).$

\subsection{Command Set}
\begin{table*}
    \centering
        \begin{tabular}{l|l|l}
            Purpose&Name & Description \\\hline
            &\texttt{setvar} & Loads a variable into the clause\\
            Setup &\texttt{validate} & Marks the clause as valid in a given context, and loads the connector bits\\
            &\texttt{chkres} & Marks the clause as valid in a given context if it is valid in any context\\
            \hline
            &\texttt{provar} & Propagates a variable in a given context\\
            &\texttt{getpro} & Gets the un-decided variable and polarity in a given context\\
            BCP &\texttt{getvar} & Get the variable at a given index\\
            &\texttt{clearvar} & Unsets a variable in a given context\\
            &\texttt{completedl} & Clears all recently-decided bits\\
            \hline
            &\texttt{copystr} & Copies the variable present bits for strengthening\\
            Strengthening &\texttt{strprovar} & Propagates a literal against the strengthening bits\\
            &\texttt{strgetpro} & Gets the propagated variable from strengthening (to delete from the learned clause)\\
            \hline
            &\texttt{clearreason} & Marks the clause as not a reason clause\\
            Learning &\texttt{getreason} & If the clause is a reason clause and the literal matches, sets the reason flag\\
            &\texttt{getlvlbits} & Gets the ``decided this level'' bits from the clause\\
            \hline
            &\texttt{nop} & No operation\\
        \end{tabular}
    \caption{The command set for a clause.}
    \label{clausecommand}
\end{table*}

The {\em decode} block in Figure~\ref{sysarch:arch} decodes
messages arriving over the network into the clause commands
listed in Table~\ref{clausecommand}.
To keep the clauses small and fast, the clause command set is extremely restricted. 

To start solving a problem, the clause must first be reset; then, the variables are loaded one at a time using the \texttt{setvar} command. The clause is next marked as valid in a single context using the \texttt{validate} command, which also loads the connector variables and latches to describe which variables are present (necessary for clause strengthening). Propagations are then sent to the clause using \texttt{provar}, which will assert the propagation flag once it has only one unset variable; the \texttt{getpro} command reads that literal and clears the flag. When a context is ready to receive clauses from another context, it can use \texttt{chkres} to do so.

Once the \texttt{getpro} command executes, it sets the reason flag, used by the clause learning and strengthening procedures. Learning starts by using the \texttt{getreason} command to search the global array for the reason clause of a given literal; once it is found, the current decision level bits are read using \texttt{getlvlbits}. This allows the clause bank controller to determine which literals need to be reasoned further and which should be sent directly to the central unit to be included in the learned clause. However, as mentioned previously, there might be duplicate reason clauses for a literal which could lead to suboptimal conflict clauses and incorrect strengthened clauses. Therefore, \texttt{clearreason} allows the central unit to send individual clauses messages telling them to ignore the fact that they are a reason clause. 

To begin clause strengthening, the central unit sends the clause unit a \texttt{strcopy} command. This loads a register with a bit for each variable that is defined; \texttt{strpro} messages are then sent for all the literals in the learned clause. These will cause the corresponding bits to be cleared in the clause unit; once all but one bit are clear \emph{and} the unset bit corresponds to a reason variable the clause will send a \texttt{strpro} message of its own. It is critical that the clause remember which bit was the reason bit, because this allows it to avoid cycles while strengthening that could result in too many variables being deleted. For example, consider a learned clause $x_0 \lor x_1 \lor x_2,$ where, on their own, $x_0$ implies $x_1,$ $x_1$ implies $x_2,$ and $x_2$ implies $x_0.$ A naive strengthening approach may incorrectly produce an empty strengthened clause.

\subsection{Connecting Variables}
Another important feature of the clause is its support for connecting variables without requiring messages to be processed by the clause bank controller. This means that each clause has connections to the previous and next clauses for propagation and strengthening. If a clause becomes unit, it marks its connecting variable as satisfied, and sends a signal to the previous or next clause to mark the corresponding connecting variable as not satisfied. This arrangement leaves open one race condition: if two adjacent clauses become unit at the same time, they may mark the same variable as decided in two different polarities! However, this will be resolved in the next cycle, because the signal from the other clause will un-satisfy the connecting variable, both clauses will assert a conflict flag, and the central unit can use the clause along with the decision trail to reason the conflict.

Because the connecting variables allow a single propagation to take arbitrarily long, an output is added to the clause that indicates whether any of the connecting variables have performed a chained propagation or strengthening. This allows the clause bank to wait the minimum amount of time before sending an idle signal, instead of waiting the worst-case 1000 cycles for a clause propagating through an entire clause bank.

\section{Methodology}
\label{sec:meth}
\subsection{Simulation}
To accurately estimate performance, we built a cycle-approximate simulator to simulate solving real SAT problems on our accelerator. The simulator is written in C++ and models three major parts:

\subsubsection{Clause Unit Behavior} 
Fixed length clause units are grouped into banks. There is a control unit inside each clause bank that bridges the interconnection network and individual clause units. The control unit decodes messages from the network into commands that are executed in parallel by all active clause units. New implications and conflicts are encoded by the control unit to become messages for the network, and message decoding and command execution are pipelined for better throughput. 
For our simulation, we use 8-variable clauses in 1024-clause banks, with a 4-stage pipelined clause bank controller.

\subsubsection{Interconnection Network Behavior}
We also simulate the behavior of the interconnection network as described in Section~\ref{sysarch:router}. Input buffering, route computation, and switch allocation are simulated at the register level to provide accurate cycle counts, including stalls from switch resource contention. Wire delay is simulated as proportional to the distance between two routers, with adjacent routers having a single-cycle delay.

\subsubsection{Central-Unit Logic}
The central unit is simulated functionally and takes care of executing the heuristics, such as decision making and restarts. Because all the computation intensive parts of solving a SAT problem are performed by the clause units, the central unit is both free from heavy workloads and flexible for configurations on different heuristics. Furthermore, because the heuristics can be executed in parallel with the accelerated computation, they are therefore modeled as incurring no extra overhead.
As previously mentioned, using even slightly different heuristics can lead to substantially different runtimes. To focus our comparison on BCP, clause learning and backtracking, we link our simulator with a \glucose solver and stay synchronized to ensure we are using exactly the same heuristics as the software solver. This is important because an implication could result in multiple different conflicts, and the conflict that is used for clause learning is then the result of a race condition. Although selecting any minimal conflict is technically correct, it makes providing a fair comparison against \glucose challenging.


\subsection{Hardware Evaluation}
Because our design is dominated by an extremely large number of clause units, we elected to optimize the design of a single clause unit and use it as a proxy for the overall design.
The design is written in parameterized Verilog HDL supporting a independently selectable context and variable counts. A series of directed tests were written to verify the core functionality of the design.
For synthesis and layout evaluation, we used the Synopsys \SI{32}{nm} educational standard cell library \cite{saed32}. 
We targeted a \SI{1}{ns} clock period at nominal voltage of \SI{1.05}{V}. 
All results are presented using the typical-typical corner at \SI{25}{C}, and all timing and power results include post place and route parasitics.

Synthesis was performed using Design Compiler, allowing full ungrouping after an initial synthesis pass and highest-effort optimizations except where otherwise stated.
After synthesis, we used IC Compiler to generate a \SI{90}{\%} utilization floorplan with a 3:1 aspect ratio, pins constrained for efficient tiling, and routing constrained to metal layers M1--M5. This allows M6--M9 to be used for routing the global clock, network links, and a global power/ground network, even though we do not route those signals in our minimal design.
After placing and routing, we used Synopsys PrimeTime to  perform a power simulation, with the post place and route netlist and back-annotated capacitances. Power simulation was performed using a \texttt{.saif} file to annotate the input and sequential elements with static probabilities and switching probabilities for a sequence of random propagations, one per cycle. We expect that this is representative of the vast majority of the SAT solver's operation. 

\section{Results}
\label{sec:res}
\subsection{Problem Characterization}
\begin{figure}
    \centering
    \begin{subfigure}[b]{0.45\textwidth}
        \centering
        \input{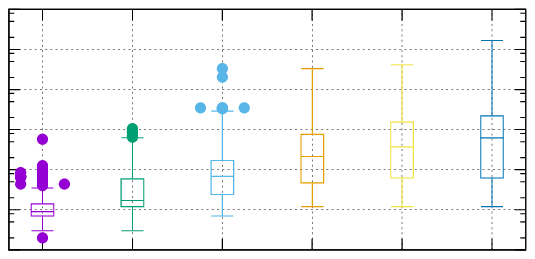}
        \caption{The number of times each variable appears.}
    \end{subfigure}
    \begin{subfigure}[b]{0.45\textwidth}
        \centering
        \input{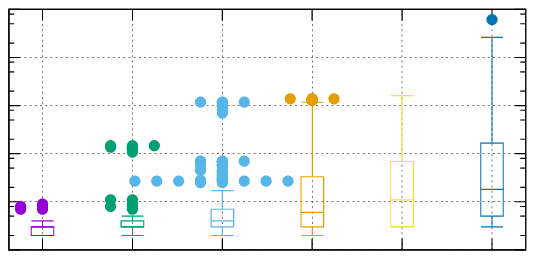}
        \caption{The number of variables per clause.}
    \end{subfigure}
    \caption{Analysis of the 2016 SAT competition application track problems without learned clauses. Each box and whisker plot shows the distribution of problems measured at a certain percentile. For example, the 99.9\% bar in (a) indicates that over all problems, the 99.9th percentile of variable popularity ranged from 10 clauses to over 10,000 clauses.}
    \label{res:nvncl}
\end{figure}

We started our search by attempting to characterize the set of interesting SAT problems. We chose the application track problems from the 2016 SAT competition as our initial benchmark set, because they are submitted from a wide variety of industrial domains \cite{sat16}. The results from this study are shown in Figure~\ref{res:nvncl} and only include the clauses present before the solver starts adding learned clauses. The majority of clauses are extremely short, with the majority of problems having 99.9\% of clauses shorter than 8 variables. They also show that the median problem has 10\% of variables in more than 10 clauses, and 1\% in almost 100 clauses. These two insights drive our design to use an associative array composed of a large number of short clauses, with each variable broadcast to the entire array. Any attempt to maintain a mapping from variables to clauses will be hampered by these popular variables.

\subsection{Software Profiling}
\begin{figure}
    \centering
    \begin{subfigure}[b]{0.45\textwidth}
        \centering
        \input{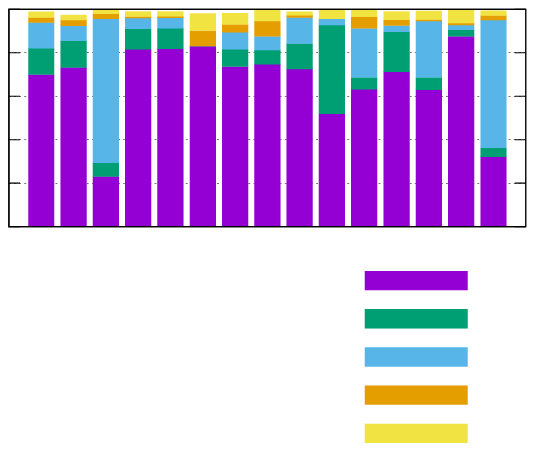}
        \caption{\minisat runtime breakdown.}
    \end{subfigure}
    \begin{subfigure}[b]{0.45\textwidth}
        \centering
        \input{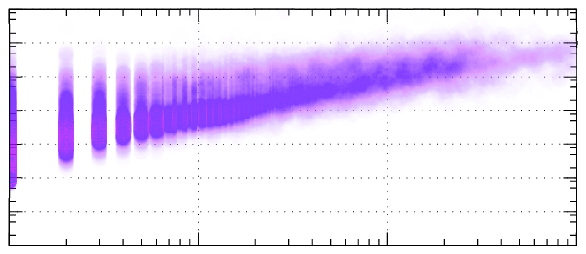}
        \caption{\glucose BCP cycles per decision level.}
    \end{subfigure}
    \caption{Software profiling results.}
    \label{res:profile}
\end{figure}

After analyzing the SAT problems themselves, we then analyzed the software solvers \glucose and \minisat, which share the majority of their code \cite{glucose, minisat}. The results of this analysis are shown in Figure~\ref{res:profile}. This profiling study highlights a key limitation of previous hardware SAT solvers, which only attempt to speed up the Boolean Constraint Propagation (BCP) step of the algorithm. Even for the problems where this would provide the most speedup, the potential speedup is still limited to 10x via Amdahl's law. For us to achieve a greater speedup, we must add in backtracking, clause learning, and clause strengthening to the hardware. Profiling also highlights the opportunities for hardware over software, with approximately 1000 cycles per implication in software and no efficient scaling for decision levels with more implications.

\subsection{Accelerator Speedup}
\begin{figure*}
    \centering
    \centering
    \input{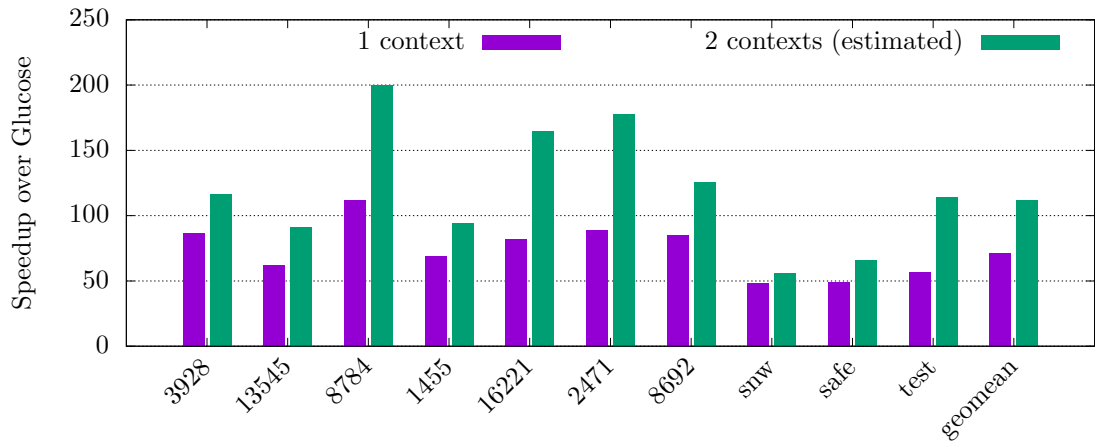}
    \caption{A comparison of \glucose against the simulated design.}
    \label{res:speedup}
\end{figure*}

We then randomly picked 10 SAT problems from 2016 SAT Competition benchmarks to solve with our simulator. The size of problems are listed in Table~\ref{res:problem}. For small problems from Agile benchmark, we solve the whole problem on  our simulator until the problem is found to be satisfiable or unsatisfiable. For large problems, with more than 300,000 clauses, simulation to completion is impractical. Instead, we terminate the simulation once we have 20,000 conflicts and compare the runtime with \glucose running to same number of conflicts (this takes about \SI{10}{s} for \glucose). Because our simulator is strictly synchronized with \glucose, the work done by the two solvers is identical. Figure~\ref{res:speedup} shows the runtime of the simulated accelerator compared to \glucose. The simulator uses a mesh topology with one execution context, a clause length of 8, and 1024 clauses per bank running at \SI{1}{GHz}. \glucose runs on a Intel i7-5930K CPU at \SI{3.5}{GHz} using a single thread.
Simulation shows that it takes 2 to 5 cycles on average for each BCP, while CPU takes around 1000 cycles. We achieve overall speedups ranging from 48x to 112x, with a geometric mean of 72x on the problems that we simulate.

\begin{table}
\centering
\begin{tabular}{l|llll}
Benchmark                    & Filename                            & \#Variables & \#Clauses \\ \hline
\multirow{7}{*}{Agile}       & bench\_3928                          & 5824        & 20353     \\
                             & bench\_13545                         & 3293        & 11661     \\
                             & bench\_8784                          & 3223        & 11057     \\
                             & bench\_1455                          & 5823        & 20338     \\
                             & bench\_16221                         & 59960       & 139918    \\
                             & bench\_2471                          & 61078       & 200509    \\
                             & bench\_8692                          & 8107        & 27257     \\ \hline
\multirow{3}{*}{Incremental} & snw\_13\_8            & 12521       & 359380    \\
                             & safe027\_pso & 74047       & 385241    \\
                             & test\_vr5\_c1\_s8257        & 89353       & 356708   
\end{tabular}
\caption{Problem characteristics}
\label{res:problem}
\end{table}

Our simulation shows that the average clause unit is 37\% idle due to network synchronization. Because different contexts synchronize independently, commands from other contexts can make full use of all clause units while one context is waiting the network to be idle. Ideally, we could speed up another 1.58x on average with multiple contexts even though we execute only one command per cycle within each clause unit. The projected speedup with two contexts is also shown in Figure~\ref{res:speedup}. 


\subsection{Network Results}
\begin{figure*}
    \centering
    \input{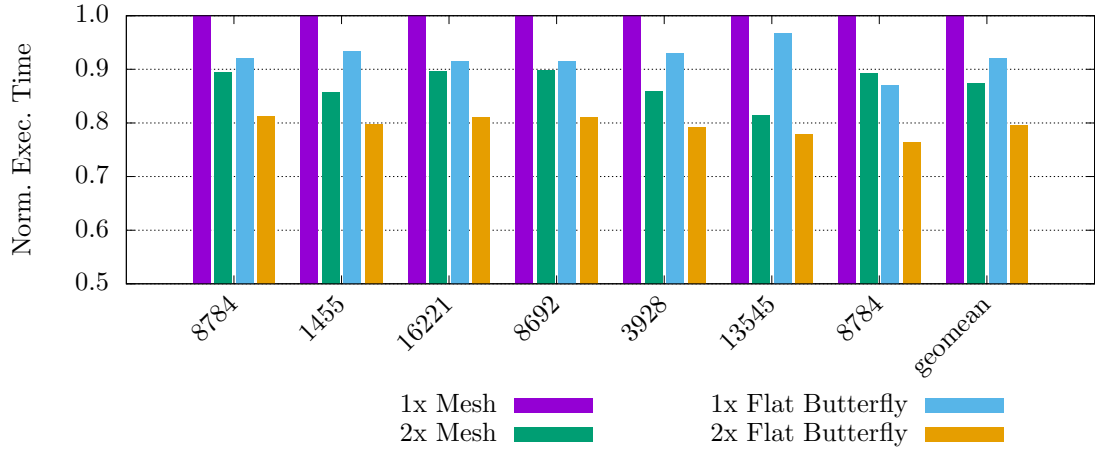}
    \caption{An evaluation of different network topologies and network/clause bandwidths.}
    \label{res:network}
\end{figure*}

We ran further experiments on our simulator to examine how the network affects performance. We simulated two topologies: mesh and flattened butterfly, and also ran the simulation with two parallel networks and two commands per clause per cycle. The results are shown in Figure~\ref{res:network} with run time normalized to the single mesh network. We get an average 12.6\% and 8.0\% speedup respectively for doubling bandwidth and switching to a flattened butterfly topology, and a 20.4\% performance improvement for both. In contrast, the increased cost for a second network and doubled clause throughput or flattened butterfly topology is much greater than its gain. For a $n \times n$ network, a flattened butterfly router needs $2n-1$ input channels and $2n-1$ output channels, while a mesh router needs only 5 channels for each direction. The cost can grow dramatically as the network gets larger, and almost all extra network bandwidth of flattened butterfly topologies is wasted by sending the same message to different routers. Therefore, we concluded it is better to use a single mesh network.
\subsubsection{Power}
We evaluate the power of the interconnection network based on two major parts: routers and wires. We implemented routers in Verilog HDL supporting both the point to point and broadcast operation, and placed and routed them using the same flow as the clause units. Each router takes \SI{22,861}{\mu\meter^2}, which is only 0.5\% of the size of one clause bank, 
and consumes almost no power.
The majority of network energy consumption comes from the interconnecting wires. With \SI{32}{nm} technology, it takes about \SI{0.2}{\pico J} per millimeter to send one bit in a densely packed bus on M7, treating the layers above and below as ground planes (the conservative case) but ignoring fringing capacitance. The overall estimation of wire power is \SI{1.85}{W} on a 16 by 16 network under the activity factor of 34\% given by the simulation. 

\subsection{Clause Layout}
\begin{figure}
    \centering
    \includegraphics[width=0.8\textwidth]{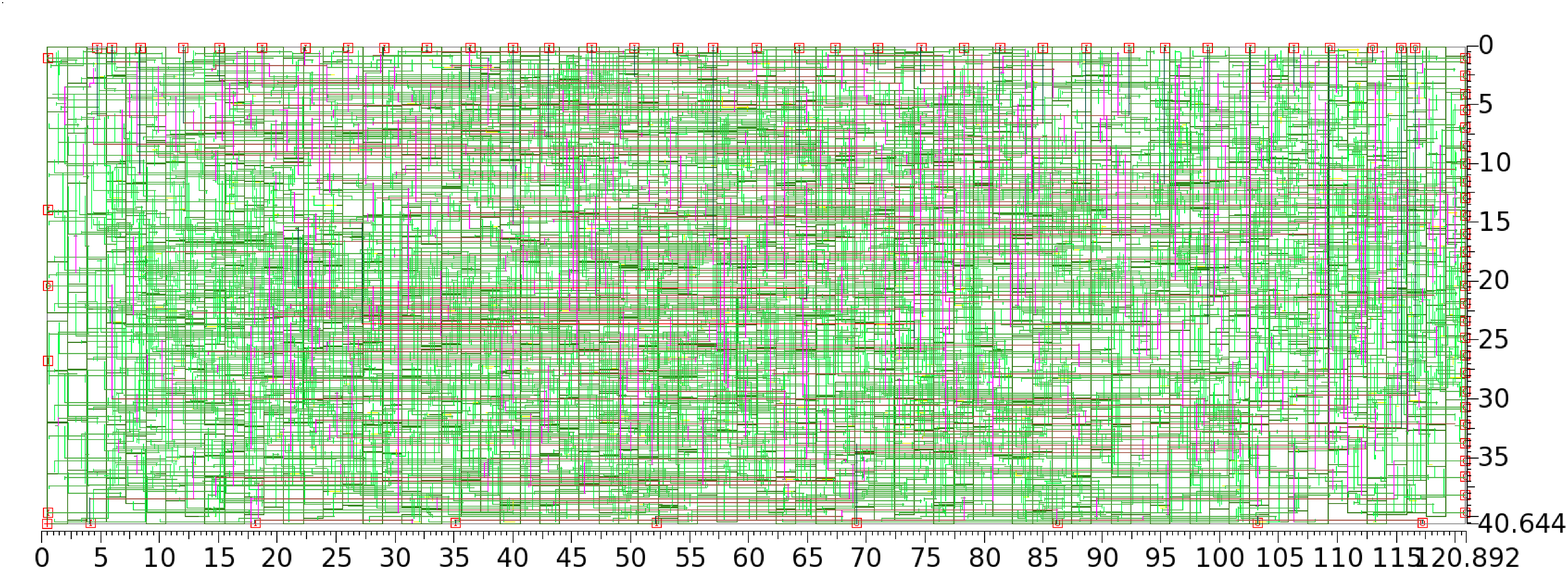}
    \caption{The placed and routed clause design, with 8 variables and 2 contexts. The design is \SI{120}{\mu m} by \SI{40}{\mu m}.}
    \label{res:layout}
\end{figure}

\subsubsection{Scaling}
Although we selected 8 variables and 2 contexts for laying out the clause unit, we also evaluated the scaling of the unit over a range of parameters (shown in Table~\ref{res:scale}). The clause unit size scales linearly with the number of variables and with the number of contexts, but there is a more significant penalty paid for the additional contexts than might be expected. This is because not all logic can be shared between contexts; notably, the individual enables for each control bit must be duplicated. Intuitively, this means that the clause units should be as small as possible because every clause unit that is too large wastes space. However, there is a practical bound on scaling because shorter clauses need more cycles to handle a propagation through connecting variables.

\begin{table}
	\centering
    \begin{tabular}{rr|l}
    Area & Power & Parameter \\\hline
    1.95x 			& 1.98x &$8\rightarrow 16$ variables \\
     1.36x          & 1.01x &$1\rightarrow 2$ contexts\\
     1.12x          & 1.33x &$20\rightarrow 24$ variable bits\\
     1.85x          & ~ &$1\rightarrow 2$ commands (old)
    \end{tabular}
    \caption{Clause scaling relative to an 8 variable, 1 context clause. Power reported is dynamic power.}
    \label{res:scale}

\end{table}

\subsubsection{Area Breakdown}
Table~\ref{res:bdown} shows the results of a synthesis flow without ungrouping to estimate the area contribution of each part of the clause architecture. We can see that sequential state makes up only a small portion of the clause, with a much larger portion being the combinational logic. The comparators themselves take around 13\% for an 8 variable clause, with an additional 8\% from a mux to select the outgoing propagation. Logic to detect the propagation and conflict states is minimal, but there is a significant control overhead. Although this cannot be broken down directly, the majority of this logic is likely in the logic to set and clear individual bits for control, as well as several smaller inferred multiplexers. The large combinational elements are heavily shared, with the output mux being used for reading out arbitrary literals (necessary for clause learning), accessing the strengthened literal, and getting propagations. Similarly, the variable finder (8 comparators, producing a one-hot output) is used to handle incoming propagations, clause learning, and clause strengthening. Other minor parts of the design include the command decoder and a shared decoder for variable and context indices (which themselves share an input bus). 

There are two types of stateful components within an individual clause. The first is the clause data---information about which variables are present, which polarities they have, etc. This is stored using D latches for space efficiency and makes up approximately 20\% percent of the total area. The other storage is replicated per context and tracks state for propagations, clause learning, and clause strengthening. This state is stored in D flip-flops.  We attempted to use D latches for this state as well, but the presence of many logical feedback loops made the implementation unworkable with our timing tools.

\begin{table}
    \centering
    \begin{tabular}{l|rr}
        Type & \multicolumn{2}{c}{Area}\\
             & ($\mu m^2$) &  (\%)\\\hline
        Clause storage & 864.1       &  20.5 \\
        Other storage & 741.6       &  17.6 \\
        Comparators (x8) &  546.7   &  13.0\\
        Variable output mux & 351.2 &  8.3  \\
        Propagation/strengthening detectors & 140.3 &      3.3  \\
        Other muxes & 273.5 &6.5\\
        Decoders & 69.6 & 1.7\\
        Other combinational area & 1230.3         & 29.2  \\
        \hline \hline
        Total cell area & 4217.3 & 100                                 \\
    \end{tabular}
    \caption{Breakdown of clause area (8 variables, 2 contexts).}
    \label{res:bdown}
\end{table}

\subsubsection{Multiple Commands}
Our current clause unit supports only a single command (of any type) per cycle. A previous design supported a parameterizable number of commands. The scaling results, also shown in Table~\ref{res:scale}, are post-synthesis only and target a smaller command set than the foregoing results, so only a relative scaling is presented. However, it is clear that increasing the number of commands increases the area linearly. This is because each additional command requires duplication of many combinational resources, especially comparators and the control logic necessary to drive the decided and satisfied bits. 

\section{Related Work}
\label{sec:rel}
There have been a number of previous attempts to accelerate SAT, but none have attained a high real-world speedup for several reasons \cite{survey2017}. 
In this section, we detail these results and explain our solver's key differences.

\subsection{Brute Force Solver}
Yuan et al. implemented a solver that uses \minisat to solve the majority of the SAT problem, then turns the remaining part over to a brute-force hardware solver \cite{yuan2012smpp}. They claim that \minisat spends the majority of its time solving the last few variables, so it is easier to turn this phase of execution over to dedicated hardware. However, their solution is limited by the exponential nature of their design: they are only able to solve for 13 variables at a time on a DE2-115 FPGA, and every additional variable would require doubling the size of the accelerator. They also do not evaluate the feasibility of their solution for the enormous problems common today.

\subsection{BCP-Only Solvers}
Another set of solvers accelerate just the BCP portion of SAT, and rely on a host CPU to perform all other functions; these have their maximum overall speedup significantly limited by Amdahl's law but are able to take advantage of advances in software heuristics.
Davis et al. implemented a BCP accelerator using FPGAs that stores the clauses in block RAMS, and obtained a speedup of 3.7--38.6x for BCP only vs. software \cite{davis2008practical}. In a large FPGA, they are able to handle 176k clauses but only 64k variables. They compress the clause representation by storing a tree mapping literals to clauses, which is walked each time an implication is processed and thus sets a bound on throughput by requiring multiple memory accesses per incoming implication. This tree representation also limits how often variables can appear, by restricting each variable to occur at most once per group of clauses. 
A further extension of this work proposes hardware to dynamically add learned clauses, but this additional hardware is not capable of learning the clauses itself \cite{davis2008designing}.

Thong et al. implemented a similar BCP accelerator, but, instead of using a tree data structure to store variables, stores them as pointers to other variables \cite{thong2013fpga}. When a variable is implied, it is traversed as a linked list with each clause pointing to the next clause containing the variable. Although this allows the clauses to be stored in block RAM, it means that these accesses must be serialized, and increases the per-implication latency. Additionally, the need to perform multiple RAM accesses per implication may limit throughput for many problems.

\subsection{Entire Problem Solvers}
Gulati et al. designed a SAT accelerator that sought to perform the entire SAT solving process in hardware \cite{gulati2009}. While their algorithm is able to perform the decisions on the accelerator (relying on the CPU to load in new sub-problems as existing ones are satisfied), they do not allow the same intelligent heuristics to be used as a software solver, and do not allow for clause learning. They estimated a speedup of 90x over \minisat, but only evaluate small problems which may not make effective use of clause learning. And as more advanced technologies like clause strengthening are not supported, it will need to search more variable assignments before reaching SAT or UNSAT.

\section{Future Work}
There are many possible future directions for this work. One is implementing a hardware version of the two-watched literal scheme implemented in software. Although this may not be efficient for the majority of clauses, it could greatly decrease the expense of keeping extremely long clauses entirely in associative storage at the cost of additional control complexity and memory bandwidth. Another potential direction is to study the performance impact of off-chip interconnection networks, and multi-chip implementations to support problem sizes larger than will fit on a single chip.

\subsection{Efficient Multiple Commands}
Using hardware to filter the commands executed by each 
clause could allow multiple commands to be executed
each cycle without requiring a linear increase in area
and power. 
Further exploring this direction would involve selecting the ideal organization of filters (one per clause vs. one per bank) and sizes and quantifying the impact via additional simulation. Another possibility includes restricting the commands that can be executed on the second command port; for example, not allowing variables to be read.

\subsection{Incomplete Algorithms}
The SAT algorithms we reference in this work are \emph{complete}---given a problem and sufficient time, they are guaranteed to produce either a satisfying assignment or a proof that the problem is not satisfiable. However, there is a class of SAT solvers that are unable to determine that a problem is not satisfiable, but may provide a satisfying assignment of variables if one exists; one such algorithm is Stochastic Local Search (SLS) \cite{sls}. SLS produces a random assignment of all variables in the problem, then greedily flips the variable that will satisfy the most clauses. It proceeds until there is either a timeout or a satisfying assignment, and is the best peforming algorithm for certain classes of SAT problems \cite{survey2017}.

\subsection{Satisfiability Modulo Theories}
Perhaps the most exciting prospects for future work lie in the field of Satisfiability Modulo Theories (SMT), an extension of SAT \cite{smt}. SMT involves specifying each variable as a logical relationship in some other theory, e.g., integers or unevaluated functions. Then, for a problem to be satisfiable, all the clauses must be satisfied \emph{and} all the logical relationships must be satisfied in the theory. The simplest SMT solvers work by translating the problem to SAT, and attempting to solve it. Once the SAT solver produces a satisfying assignment, the theory solver attempts to determine if it is logically true. If not, a clauses is added to the SAT problem disallowing it and the search continues; otherwise, the search stops. 


\section{Conclusion}
\label{sec:concl}
In this work, we introduced SatIn, an accelerator for boolean satisfiability that uses a distributed control scheme and is partitioned over a network. This is different from many current accelerators, which do not support spatial partitioning for solving arbitrarily large problems. We demonstrated that our individual computing elements (clauses) consume a small amount of area and power and can be tiled into large arrays. Using 8 variable, 2 context clauses we can fit a single clause into \SI{4867}{\mu m^2} consuming \SI{63.8}{\mu W} dynamic power in \SI{32}{nm}. This is sufficient, along with a reasonable scaling factor, to solve problems composed of hundreds of thousands of clauses on a single chip in a more modern process.
    We simulated SatIn with a cycle-accurate simulator on real benchmarks and achieved an average 72x speedup against \glucose, the winner of the 2016 SAT competition, using a single context. 

\bibliographystyle{IEEEtran}
\bibliography{main}
%
%
%


\end{document}